\documentclass[12pt]{article}
\pdfoutput=1
\usepackage{geometry,enumerate,amsmath,amssymb}
\usepackage{fullpage}
\usepackage{graphicx}
\usepackage{bm}
\newcommand{\be}{\begin{equation}}
\newcommand{\ee}{\end{equation}}
\newcommand{\bea}{\begin{eqnarray}}
\newcommand{\eea}{\end{eqnarray}}

\newcommand{\bnabla}{\mbox{\boldmath $\nabla$}}
\renewcommand{\epsilon}{\varepsilon}
\usepackage{booktabs}
\numberwithin{equation}{section}

\newcommand{\reffig}[1]{Fig.~\ref{#1}}

\begin{document}
\title{Dynamics of linked filaments in excitable media}

\author{
  Fabian Maucher$^{\dagger\star}$ and Paul Sutcliffe$^\star$\\[10pt]
{\em \normalsize
$^\dagger$Joint Quantum Centre (JQC) Durham-Newcastle, Department of Physics,}
  \\
{\em \normalsize  Durham University, Durham DH1 3LE, United Kingdom.}\\
 {\em \normalsize $^\star$Department of Mathematical Sciences,}\\
 {\em \normalsize Durham University, Durham DH1 3LE, United Kingdom.}\\ 
{\normalsize Email: fabian.maucher@durham.ac.uk, \ p.m.sutcliffe@durham.ac.uk}
}
\date{April 2018}

\maketitle

\begin{abstract}
  In this paper we present the results of parallel numerical computations of the long-term dynamics of linked vortex filaments in a three-dimensional FitzHugh-Nagumo excitable medium. In particular, we study all torus links with no more than 12 crossings and identify a timescale over which the dynamics is regular in the sense that each link is well-described by a spinning rigid conformation of fixed size that propagates at constant speed along the axis of rotation. We compute the properties of these links and demonstrate that they have a simple dependence on the crossing number of the link for a fixed number of link components. Furthermore, we find that instabilities that exist over longer timescales in the bulk can be removed by boundary interactions that yield stable torus links which settle snugly at the medium boundary. The Borromean rings are used as an example of a non-torus link to demonstrate both the irregular tumbling dynamics that arises in the bulk and its suppression by a tight confining medium. Finally, we investigate the collision of torus links and reveal that this produces a complicated wrestling motion where one torus link can eventually dominate over the other by pushing it into the boundary of the medium. 
\end{abstract}

\newpage
\section{Introduction}\quad
Ever since the pioneering work of Lord Kelvin \cite{Kelvin}, there has been a fascination with the study of knotted and linked vortex filaments.
Recent research on this topic covers many areas including
hydrodynamics \cite{KI}, optics \cite{Den}, liquid crystals \cite{AS2}, relativistic field theory \cite{FN}, atomic condensates \cite{Bar} and ferromagnetic materials \cite{Sut}. The host system of interest in the present paper is that of an excitable medium, specifically the FitzHugh-Nagumo medium \cite{FH,Nag}, that is used to model electrical waves in cardiac tissue \cite{Kog} and neurons \cite{Izh}. 

The FitzHugh-Nagumo equation is a nonlinear partial differential equation of reaction-diffusion type, so the evolution of
spiral wave vortex filaments takes place within the realm of non-equilibrium dynamics. It is therefore perhaps surprising that several decades ago it was hypothesized that knotted or linked vortex filaments in this excitable medium might preserve topology, as a result of short-range repulsive interactions between filaments \cite{Winfree:Nature:1984}.
Numerical evidence for this conjecture has been found by direct long-term evolution \cite{Sutcliffe:PRE:2003} and 
by complex untangling dynamics of vortex filaments without untying \cite{Sutcliffe:PRL:2016}.

The preservation of topology alone is not sufficient to conclude that a steady regular conformation of the filament emerges at late times. Indeed the generic case for knots is a permanently irregular (though topology-preserving) evolution in the bulk of the medium, but it has been found that torus knot filaments can be stabilized by their interaction with the medium boundary, where they settle to particular conformations that appear to be stable attractors of the confined system \cite{Sutcliffe:PRE:2017}. The twist of a filament, which is an automatic consequence of filament linking or threading \cite{winfree:physD:1985}, plays a crucial role in the evolution.
It has been shown that linked vortex rings can have enhanced stability compared 
to their unlinked counterparts, with the linking inherently yielding a non-vanishing total twist of the ring \cite{Winfree:SIAM:1990, Winfree:Nature:1994}. 
Furthermore, twist can mediate long-range interactions between vortex rings \cite{Sutcliffe:JPhysA:2018}. This suggests that there might be differences in the dynamics of knots and links due to the fact that local twist can only be redistributed around each component of a link and cannot be shifted between different components. 

In this paper, we systematically study the dynamics of filament links, including all torus links with no more than 12 crossings. First, we show that all these torus links can be stabilized by the boundary of the medium, in analogy to the situation for torus knots. We introduce new definitions of position and size for a filament link in an excitable medium via moments of a vorticity. Not only is our definition of size easier to compute than the length of a link, but we observe that torus links acquire a minimal size that has a simple dependence on the crossing number of the link and varies with the number of components. For torus links evolving in the bulk, we find that there is a time interval over which the filament motion is regular, in the sense that the size, velocity and rotation frequency all remain approximately constant (when averaged over a spiral vortex period). We find a simple fit for these quantities as a function of crossing number, but the result depends upon the number of components of the link.  

The evolution of non-torus links is much more involved. The generic outcome is a complex irregular tumbling dynamics for filaments that typically leads to filaments breaking at the boundary of the medium.
However, there are examples where a confined medium can replace this type of dynamics by a regular motion that remains close to the medium boundary, as demonstrated by the example of the Borromean rings. Finally, we turn our attention to the collision of a pair of initially well-separated torus links. We consider the simplest case of a pair of Hopf links and show that if the impact parameter is zero then the pair settle into a persistent mutual locked configuration but otherwise the evolution is irregular as the pair wrestle and tumble around each other, leading to a variety of different final outcomes. The collision of a pair of links with different crossing numbers produces a wrestling between two unequal opponents and the link with the smallest crossing number can eventually dominate and push the other link into the boundary of the medium.

\section{The FitzHugh-Nagumo equation and filaments}\label{sec:FN}\quad
The FitzHugh-Nagumo equation \cite{FH,Nag} is the archetype of a mathematical model of an excitable medium and is used as an action potential description of cardiac dynamics \cite{Kog} and neuron activity \cite{Izh}. It is a two-component reaction diffusion equation given by
\begin{align}
\frac{\partial u}{\partial t}&=\frac{1}{\epsilon}(u-\frac{1}{3}u^3-v)+\nabla^2 u,\\
\quad
\frac{\partial v}{\partial t}&=\epsilon(u+\beta-\gamma v),
\label{eq:FHN}
\end{align}
where $u({\bf r},t)$, represents the electric potential and 
$v({\bf r},t)$ is the recovery variable, both being real-valued physical fields defined throughout the three-dimensional medium with spatial coordinate ${\bf r}=(x,y,z)$ and time $t$. The constants $\epsilon,\beta,\gamma$ control the properties of the medium and the behaviour of the vortex filaments that it hosts. In this paper we choose the parameter set $\epsilon=0.3, \ \beta=0.7, \ \gamma=0.5$,
to ensure a positive filament tension for untwisted filaments with small curvature and to eliminate vortex meander \cite{Winfree}.

A two-dimensional medium with these parameter values supports a rotating spiral wave vortex \cite{Winfree} with a period $T=11.2$ and $u$ and $v$ wavefronts in the form of an involute spiral with a wavelength $\lambda=21.3$. The typical dimensions of our medium will be of the order of several wavelengths and when we refer to long-term dynamics we mean timescales of hundreds to thousands of periods. A vortex filament is a structure in a three-dimensional medium whose local two-dimensional cross-section normal to the filament contains a spiral wave vortex. A straight filament is the simplest example, in which the two-dimensional vortex is embedded by assuming independence in the third dimension, but then the filament must end on the boundary of the medium. The most elementary example of a vortex filament that is contained entirely within the medium is a vortex ring, where the filament is a circle. We shall briefly review this well-studied case in the following section, to introduce some useful concepts, before moving on to our main topic of multiple linked filaments. 

To analyze and visualize filaments it is helpful to introduce the quantity 
\begin{equation}
 {\bm B} = \bnabla u \times \bnabla v,
\end{equation}
which we refer to as vorticity, because $|{\bm B}|$ is highly localized
around the vortex filament due to the fact that
in a two-dimensional vortex it
is maximal at the centre of the vortex \cite{Win3}.
Filaments are visualized by plotting tubes given by the isosurface $|{\bm B}|=0.1$, so that the centrelines of these tubes are the vortex filaments.
We illustrate filament twist, as in \cite{Sutcliffe:JPhysA:2018}, by colouring the isosurface by the value of the phase
\begin{equation}
 \varphi = \tan^{-1}\left(\frac{2(v+0.4)}{u+0.4}\right).
 \label{eq:angle}
\end{equation}

For filament links (or knots) we define the centre of mass and a concept of size by calculating moments of the vorticity, as follows. For any vector (or scalar) field ${\bf w}$, given throughout the medium, we define the filament averaged value to be
\be
\langle{\bf w}\rangle=\bigg(\int_\Omega {\bf w} |{\bm B}|^4 \,d^3r\bigg)
\bigg(\int_\Omega |{\bm B}|^4 \,d^3r\bigg)^{-1},
\label{average}
\ee
where $\Omega$ is the domain of the medium. The position of a link is defined to be ${\bf r}_o=\langle {\bf r} \rangle$ and hence its velocity is
$\dot{\bf r}_o$, where dot denotes differentiation with respect to time.
As all the links we consider have a non-zero velocity,
we can define the projector ${\cal P}$, that projects any vector field ${\bf w}$ onto the plane perpendicular to the link velocity, by
\be
   {\cal P}{\bf w}={\bf w}-\frac{\dot {\bf r}_o}{|\dot {\bf r}_o|^2}({\bf w}\cdot \dot {\bf r}_o).
   \ee
   We  define the size of a link, $s$, as a measure of the transverse extent of the link, using the formula
   \be
   s=\sqrt{\langle |{\cal P}({\bf r}-{\bf r}_o)|^2\rangle}.
   \ee
   In the simple case that the filament is a vortex ring, the size should be a good approximation to the radius of the ring, and indeed it is, as we demonstrate in the next section. In fact this requirement is the motivation for choosing the quartic average in (\ref{average}) rather than the more obvious quadratic average. Not only is the size easier to compute than the length of a link, but it will turn out that it correlates very well with the crossing number of the link, thereby providing a useful connection between topology and conformation.
   
 An important aspect of filament dynamics is that filaments are
 continuously emitting wavefronts that can slap filaments and drive their motion \cite{Winfree:book:2002}. Some insight into this phenomenon can be found by considering the time evolution of the vorticity,
\be
  \frac{\partial {\bm B}}{\partial t}=\frac{(1-\epsilon^2\gamma-u^2)}{\epsilon}{\bm B}+(\nabla^2 \bnabla u)\times\bnabla v.
   \label{eq:B}
\ee
Ignoring the higher derivative term, we see that the interaction between the
excitation wavefront in $u$ and the vorticity is generated by the nonlinear term  proportional to $u^2{\bm B}$.
Note that, in principle, the range of this interaction is limited only by the size of the medium. However, as excitation waves annihilate on collision with each other, in practice the range of the interaction depends in a complicated manner of the location of the collision interface. In the simplest case of a pair of parallel straight filaments that are close together, this interaction typically gives rise to the separation of the filaments to a certain distance, at which the the filaments then remain. However, even in the simplest linked case of vortex rings threaded by straight filaments, it has been found that twist mediated interactions can produce complicated dynamics that can even reverse the direction of motion of a ring \cite{Sutcliffe:JPhysA:2018}.
The dynamics of linked filaments, with multiple twisted components, is expected to be even more complicated.

\begin{figure}[h!]
\begin{center}
 \includegraphics[width=1.0\columnwidth]{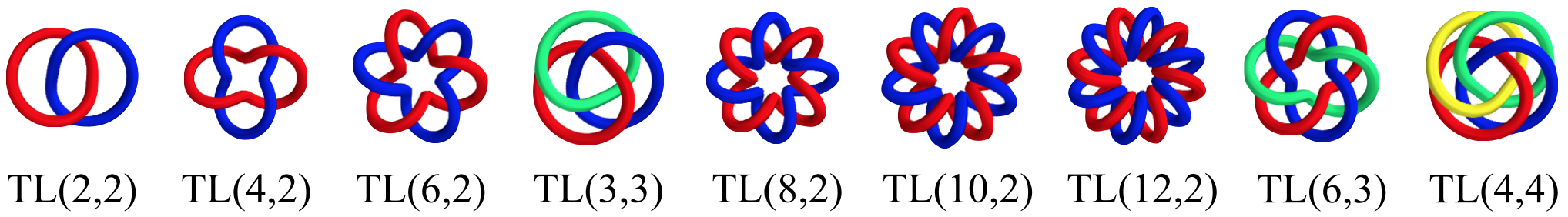}
\end{center}
 \caption{The 9 torus links with crossing number no greater than 12. The links are ordered by crossing number and the colours distinguish the different components of the link.
 }
\label{fig:TL}
\end{figure}

Torus links are classified by a pair of integers $p\ge q>1$ that are not coprime. We denote such a torus link by the notation $\mathrm{TL}(p,q)$, where the two integers count the number of times that the link winds around the poloidal and toroidal directions of the torus on which it may be inscribed. The (minimal) crossing number of the link is $N=p(q-1)$ and the number of components is the greatest common divisor of $p$ and $q$. Although a knot is technically a one-component link, in this paper we shall reserve the use of the term link to refer to only those with multiple components, in order to distinguish filament links from the filament knots that have been studied previously \cite{Sutcliffe:PRE:2017}.
There are 9 torus links with crossing number $N\le 12$ and these are depicted
in~\reffig{fig:TL}, where they are ordered by increasing crossing number and colours are used to distinguish the different components of a given link. The link with two crossings, $\mathrm{TL}(2,2)$, is known as the Hopf link and the link with four crossings, $\mathrm{TL}(4,2)$, is confusingly known as Solomon's knot, despite the fact that it is a link and not a knot.

Initial conditions for torus links in the FitzHugh-Nagumo equation are obtained using Milnor maps \cite{Milnor},
as described in \cite{Sutcliffe:PRE:2003}.
Explicitly, the spatial coordinate ${\bf r}=(x,y,z)$ is mapped to the complex coordinates $(\eta,\zeta)$, that lie on a unit three-sphere in ${\mathbb C}^2$, via
\be
(\eta,\zeta)=\frac{1}{x^2+y^2+z^2+L^2}\big(2L(x+iy),x^2+y^2+z^2-L^2+2iLz\big),
\ee
where the real constant $L$ is positive and simply sets the length scale of the initial link. Initial fields for the torus link $\mathrm{TL}(p,q)$ are given by
\be
u({\bf r},0)=2\Re(\eta^p+\zeta^q)-0.4,\qquad\qquad
v({\bf r},0)=\Im(\eta^p+\zeta^q)-0.4,
\ee
where $\Re$ and $\Im$ denote the real and imaginary parts.
This creates a torus link with an initial position on the $z$-axis, ${\bf r}_o=(0,0,z_o)$, and an initial velocity aligned with this axis, $\dot{\bf r}_o=(0,0,\dot z_o)$. 

To provide initial conditions for any non-torus link we apply the method introduced in \cite{Sutcliffe:PRL:2016}. This involves specifying discrete linear segments for each component of the link and computing the vector field defined by a
Biot-Savart integral along the discretized link. The initial fields $u({\bf r},0)$ and $v({\bf r},0)$ are then obtained from the value of the scalar potential of the vector field, calculated as a line integral. Note that the same method can also be used to provide initial conditions for torus links, but the approach using Milnor maps is more convenient.

We employ parallel numerical computations to calculate solutions of the FitzHugh-Nagumo equation (\ref{eq:FHN}) using standard methods, with time evolution performed using a fourth-order Runge-Kutta method with timestep $\Delta t=0.1$ and spatial derivatives calculated using the discrete cosine transform with a lattice spacing $\Delta x=0.5$. The shape of the medium is a cuboid, with faces normal to the coordinate axes, on which no-flux (Neumann) boundary conditions are imposed.

\section{Dynamics of torus links}\quad
Before we turn our attention to torus links, we first illustrate our use of vorticity moments to determine size and position by considering the well-studied case of a single vortex ring.
A motivation for introducing our new definition of the size of a filament link is to provide a simple method to interpret the data, without the need to resort to more involved computations, such as computing the length of a filament via a discrete Fourier transform and interpolation \cite{Sutcliffe:PRE:2017}.
\begin{figure}[ht!]
\begin{center}
 \includegraphics[width=0.6\columnwidth]{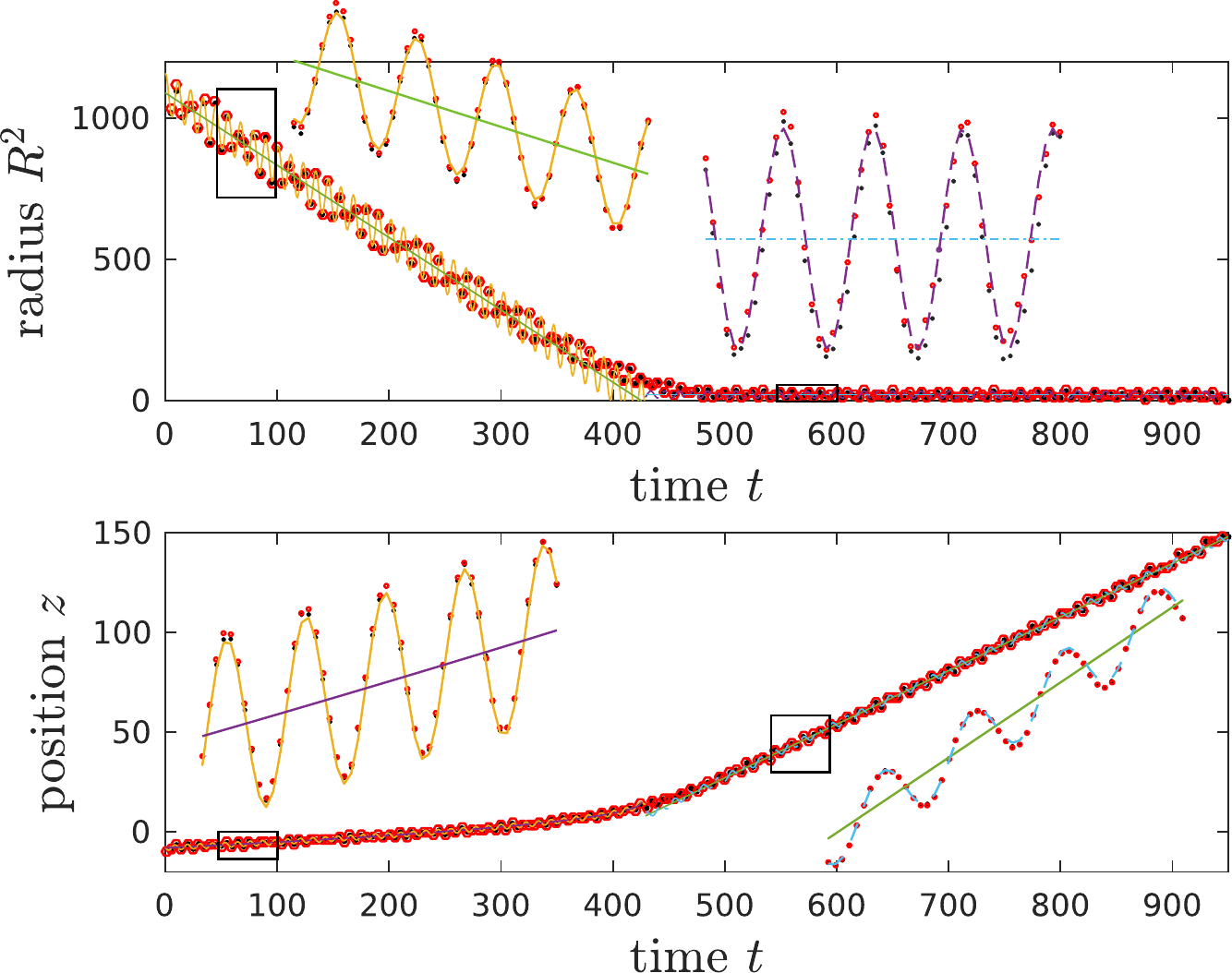}
\end{center}
 \caption{The dynamics of the square of the radius (upper image) and the position along the symmetry axis (lower image) of a single vortex ring.  
   The black dots represent numerical data computed using the more traditional method of interpolation and the red circles use the new method of vorticity moments. It can be seen that the results from these two methods agree and match very well to the oscillating curves that are fits to the numerical data.
The lines are the analytic approximations for the period-averaged values. There are two regimes, one where the 
 radius shrinks and one where the radius remains constant. The speed of the ring is constant in the regime where the radius is constant. Insets show magnified regions of the data and fits. 
  }
\label{fig:unknots}
\end{figure}

 \begin{figure}[h!]
  \begin{center}
 \includegraphics[width=0.9\columnwidth]{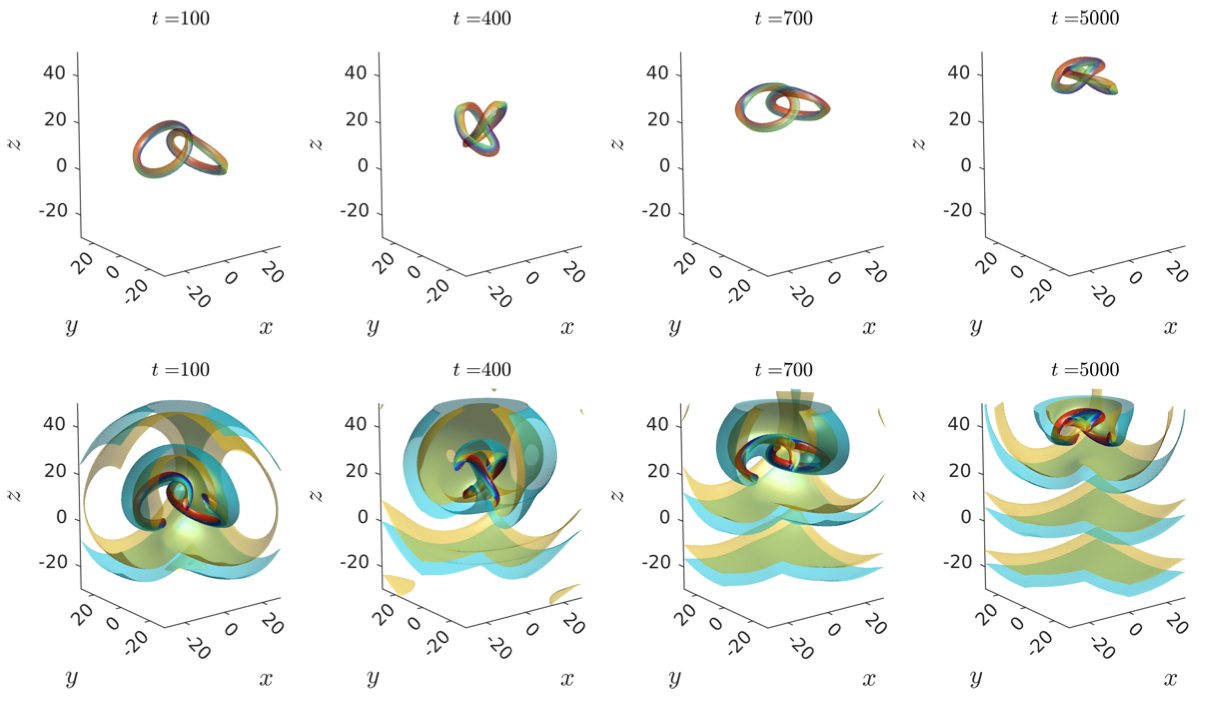}
 \caption{The dynamics of a Hopf link. The upper images display the evolution of the filament of the link, coloured by the phase $\varphi$ to illustrate twist.
The lower images are reproductions of the upper images but also show wavefronts given by $u=0$ isosurfaces coloured by the value of the phase, with a quarter of each plot removed to aid visualization.  
The waves emitted from both components of the link yields a mutual slapping
and the overall result is a rotation and translation of the link.
 }
 \label{fig:Hopf_self_propelled}
 \end{center}
 \end{figure}
 \begin{figure}[h!]
\begin{center}
 \includegraphics[width=1.0\columnwidth]{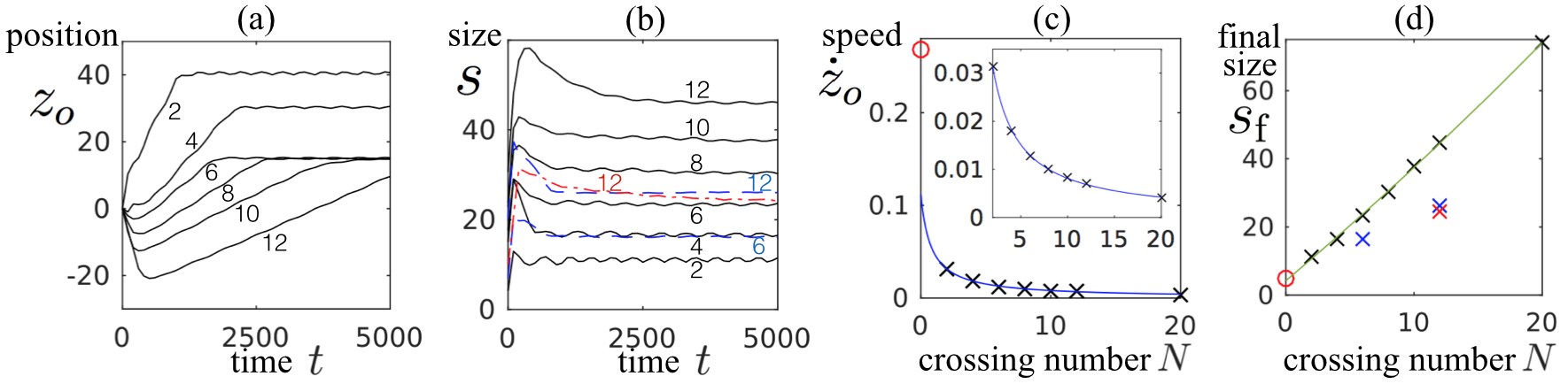}
\end{center}
\caption{The size and motion of torus links. (a) the position $z_o$ along the symmetry axis as a function of time for all 6 of the 2-component torus links with crossing number $N\le 12.$ \quad
  (b) the size $s$ as a function of time for all 9 torus links with crossing number $N\le 12.$ Black, blue and red curves denote links with 2, 3 and 4-components.
  (c) the speed $\dot z_o$, during regular bulk motion, as a function of crossing number $N$ for all 6 of the 2-component torus links with $N\le 12$, plus $N=20$ (black crosses) and the vortex ring (red circle). The curve is the fit (\ref{fitspeed}).
  (d) The final size of the links at the medium boundary as a function of crossing number $N$. Black crosses correspond to all 6 of the 2-component torus links with $N\le 12$ plus $N=20$. Blue and red crosses denote torus links with 3 and 4 components and the red circle is the vortex ring. The curve is the fit (\ref{fitsize}) for the 2-component links.
  }
\label{fig:torus_links}
 \end{figure}
\begin{figure}[h!]
\begin{center}
 \includegraphics[width=0.7\columnwidth]{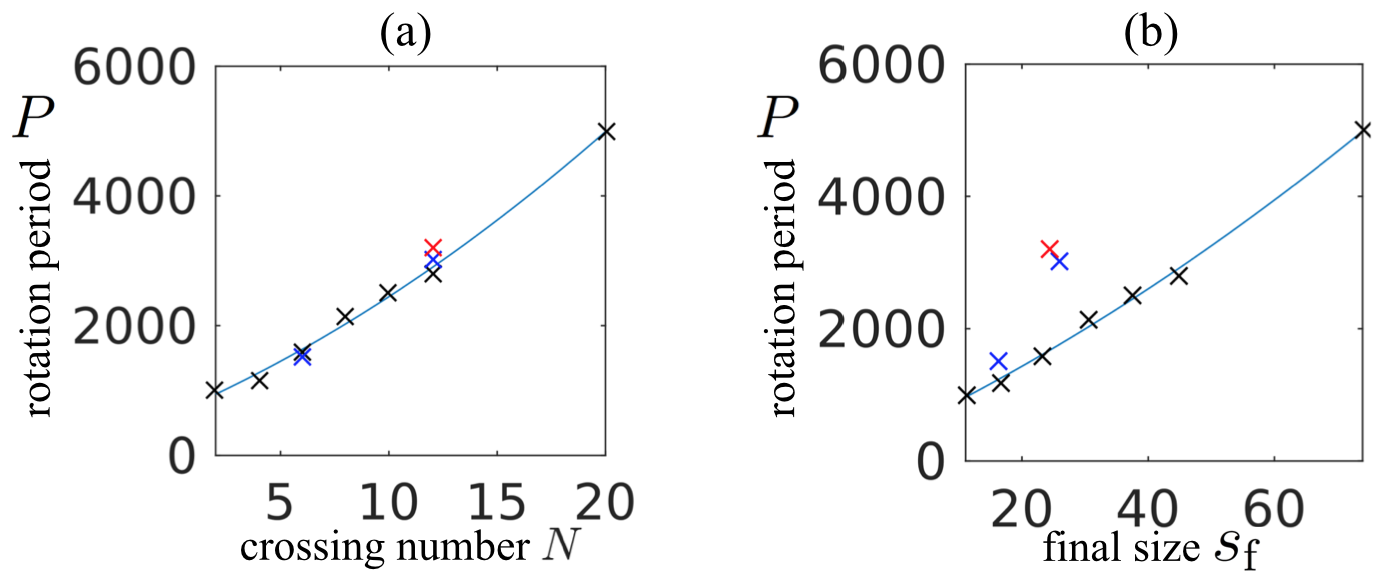}
\end{center}
 \caption{The rotation period for torus links at the medium boundary, (a) as a function of the crossing number $N$, (b) as a function of the final size $s_{\rm f}$ of the link. Black, blue and red crosses correspond to links with 2, 3 and 4 components respectively. The curve in (a) is the fit (\ref{fitperiod}) and the the curve in (b) is obtained by combining the fit (\ref{fitperiod}) with the fit (\ref{fitsize}). 
 }
\label{fig:period}
\end{figure}

It is been derived analytically and demonstrated numerically that in the case of positive filament tension the period-averaged radius $R(t)$ of an initially large circular vortex ring shrinks according to the formula $R^2(t)=R^2(0)-2\tau t$,
where the positive constant $\tau$ is by definition the filament tension  
\cite{Keener:PhysicaD:1988,Biktashev:1994}. As it shrinks, the vortex ring propagates along its symmetry axis. Once the ring attains its minimal radius the evolution enters a second regime where both the
period-averaged radius and the period-averaged speed remain constant.
The black dots in \reffig{fig:unknots} reproduce the data from \cite{Sutcliffe:JPhysA:2018}, that uses a standard interpolation method to locate the filament and compute its position and length. The oscillating curves are numerical fits to this data and the lines are the analytic approximations for the period-averaged values. The vortex ring is initialized so that the $z$-axis is the axis of symmetry and the position refers to the centre of mass along this axis.  The red circles in \reffig{fig:unknots} denote the position ${\bf r}_o=(0,0,z_o)$ and square of the size $s^2$ computed using the definitions in the previous section from the moments of the vorticity. It is clear that the new method produces results that are in excellent agreement with the more traditional computation. Given the ease of calculating vorticity moments, we shall use this method in our analysis of links. As mentioned earlier, the excellent agreement between the results for the two methods is the motivation for choosing the quartic average in (\ref{average}).

 Using the initial conditions described in the previous section, we make a systematic study of the dynamics of all torus links with crossing number no greater than 12. The simplest example, the Hopf link, is presented in \reffig{fig:Hopf_self_propelled}, where the upper images display the evolution of the link filament and the lower images also include wavefronts given by $u=0$ isosurfaces.
 Filaments and wavefronts are coloured according to the value of the phase $\varphi$ in order to illustrate twist. This colouring confirms that each component of the link is twisted through one total revolution, in agreement with the general theorem \cite{winfree:physD:1985} that for each component of a link the total twist (defined with an appropriate sign convention) is equal to the sum of its linking numbers with all components. The twisting of each filament results in a time lag for wave emission at different points along the filament and the overall motion of the link is a translation along the $z$-axis accompanied by a rotation around this axis. 

We have found a similar evolution for all torus links with crossing number up to 12, with all displaying a rigid rotation and translation with a well-defined minimal size, in analogy to previous results on torus knots \cite{Sutcliffe:PRE:2017}. A priori, the same behaviour for knots and links is not guaranteed because links have additional degrees of freedom. In particular, an important difference for links is that once a component develops a local excess of twist, other components cannot participate in reducing and redistributing this twist. This contrasts with the situation for a knot, where the whole length of the filament 
can participate in reducing the local twist rate to a minimum.

The results presented in \reffig{fig:torus_links} display detailed information about the motion and size of torus links and how these properties relate to the crossing number $N$ of the link. In \reffig{fig:torus_links}(a) the position $z_o$ along the $z$-axis is shown as a function of time for all 2-component links with $N\le 12$. Similar data has been obtained for the torus links with 3 and 4 components but for clarity these results are not shown on this figure. The dynamics can be described by three different stages. For each link there is an initial transient stage as the filaments of the link are formed and the link can move either up or down the axis in a manner that depends on the details of the initial condition. In the second stage the link has attained its minimal size and travels within the bulk of the medium at an approximately constant speed $\dot z_o$. This stage of the motion emulates the behaviour found in the second regime of motion for a vortex ring, where the ring travels at a constant speed once it has acquired its minimal size. In the third stage the link has settled snugly at the boundary of the medium, given by $z=50$ for $N=2$, $z=40$ for $N=4$, and $z=25$ for $N\ge 6$, to reflect the different speeds of these links. Note that the torus link $\mathrm{TL}(12,2)$ has not quite reached the boundary in the time displayed in \reffig{fig:torus_links}(a). This third stage of the motion has no analogue for the vortex ring, as the ring annihilates with its mirror image on reaching the boundary of the medium.  

The evolution of the size $s$ of the link is shown in \reffig{fig:torus_links}(b) for all 9 torus links with $N\le 12,$ with colour coding for the number of components of the link. After the transient first stage, there is a rapid decrease in size to an asymptotic minimal size, $s_{\rm f}$, and this final size is maintained when the link settles at the boundary in the third stage of the motion. The speed during the second phase of the motion is plotted as a function of crossing number $N$ in \reffig{fig:torus_links}(c) for all 6 of the 2-component torus links with $N\le 12$, plus $N=20$ (black crosses) and the vortex ring (red circle).
The curve is a reciprocal linear fit to the data given by
\be
\dot z_o=\frac{0.088}{N+0.785}.
\label{fitspeed}
\ee
Note that setting $N=0$ in the above formula does not provide a good approximation for the speed of the vortex ring, suggesting that this formula only applies to 2-component torus links. 

The final size $s_{\rm f}$ of each torus link, computed once it has settled at the boundary, is plotted in \reffig{fig:torus_links}(d), as a function of the crossing number $N$. The crosses are colour coded, black for 2-component links, blue for 3-component links and red for the 4-component link. The red circle is the size of the vortex ring. The final size of 2-component torus links displays an almost linear dependence on the crossing number, with the optimal quadratic fit to the data given by
\be
s_{\rm f}=4.249+3.151\,N+0.017\,N^2,
\label{fitsize}
\ee
where we note the small coefficient of the quadratic term. This fit is included as the curve in \reffig{fig:torus_links}(d), where we see that it provides an excellent estimate for the final size of 2-component links, but is not applicable to links with more than 2 components. Setting $N=0$ in formula (\ref{fitsize}) does provide a reasonable estimate for the size of the vortex ring, suggesting that the first term in (\ref{fitsize}) might be universal but that the remaining coefficients depend on the number of components of the link.

Once a torus link settles at the boundary of the medium then obviously its speed is zero, however, the link continues to rotate about an axis normal to the boundary and therefore we can compute the period $P$ of this rotation. The results of these calculations are shown by the crosses in \reffig{fig:period}, where we use the same colour coded scheme as earlier to indicate the number of components of the link.  A quadratic fit to the data for 2-component links gives the following approximate formula for the period $P$ as a function of crossing number $N$
\be
P= 643+143\,N+3.7\,N^2.
\label{fitperiod}
\ee
In \reffig{fig:period}(a) it can be seen that this approximation also provides a reasonable estimate of the period of the 3-component and 4-component links with $N\le 12$, suggesting that the rotation period may have a universal behaviour that is independent of the number of components of the link. As shown earlier, the size of the link does depend on the number of components, so this implies that there cannot be a component-independent relation between the rotation period and the final size of the link. This is confirmed by the plot in \reffig{fig:period}(b), where the rotation period is shown as a function of the final size and the curve obtained by combining the expressions (\ref{fitsize}) and (\ref{fitperiod}) provides a good estimate for 2-component links, but not for links with more components.

As the rotation period increases with crossing number $N$, we expect that the rotation frequency tends to zero in the asymptotic limit of large $N$.
This can be understood intuitively by observing the dynamics with a fine time resolution, where it becomes clear that the rotation is not smooth but rather can be described as a two steps forward, one step back type of motion, similar to the escapement mechanism of a clock. In other words, the link first rotates in one direction and then rotates in the opposite direction by a smaller amount.
As the crossing number increases the forward and backward steps become more comparable, presumably because there is less difference between the inner and outer parts of the filament as it winds around the torus, and the result is a decrease in the net rotation frequency. 

\section{Dynamics of the Borromean rings}\quad
The generic dynamics for links that begin their evolution in the bulk of a medium is an irregular tumbling dynamics that preserves topology but not conformation and eventually leads to filaments breaking at the boundary of the medium. The regular dynamics of torus links presented in the previous section is exceptional behaviour and mirrors that found previously for torus knots \cite{Sutcliffe:PRE:2017}. For both torus knots and torus links, sufficiently symmetric initial conditions produce translational motion and if the knot or link can reach the boundary of the medium before any instabilities have developed then a boundary stabilized configuration with a periodic evolution is obtained.
\begin{figure}[h!]
\begin{center}
 \includegraphics[width=0.9\columnwidth]{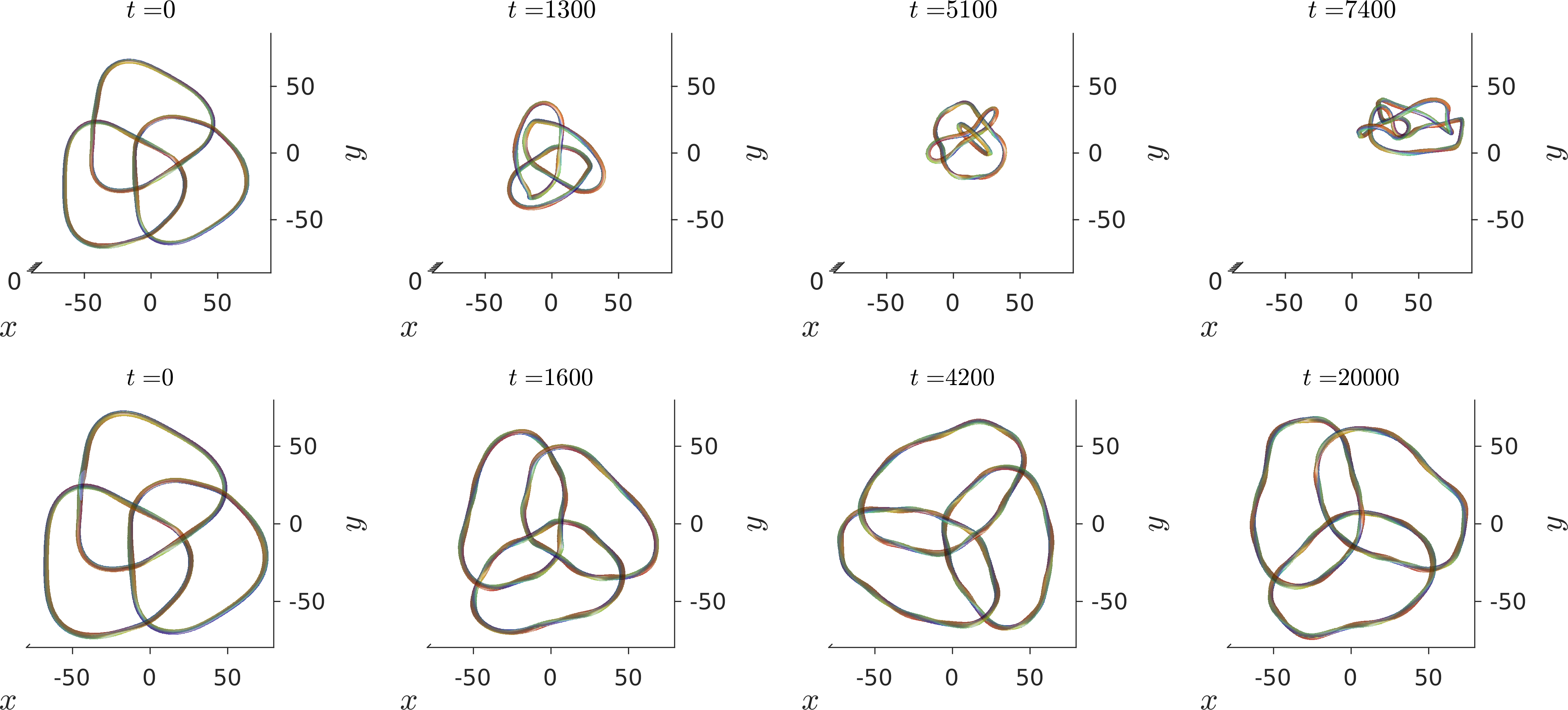}
\end{center}
\caption{The same symmetric initial condition for Borromean rings leads to irregular tumbling dynamics in a sufficiently large medium (upper panel) but regular dynamics near the boundary of a more tightly confined medium (lower panel). Note that the total simulation time is much greater for the lower panel and that the scale is slightly different in the two panels because the medium in the upper panel is slightly larger. Eventually the tumbling dynamics leads to the filament breaking (not shown) at the boundary of the medium. 
 }
\label{fig:Borromean}
\end{figure}

As we have seen, as the crossing number of a torus link increases then its speed decreases, hence it takes longer to reach the medium boundary and this gives more time for any instability to develop. The computations of the speed of torus links presented in the previous section require a sufficient period of regular
translational motion within the bulk. Even for the torus links with crossing number no greater than 12, we found that the link $\rm{TL}(4,4)$  did not travel a sufficient distance during regular motion to be able to reliably assign a speed. However, links with this type of behaviour can be stabilized by creating the link sufficiently close to the boundary of the medium and this allows the size of the link to be computed.

A striking demonstration of the difference between bulk evolution and boundary stabilized dynamics is provided by considering the Borromean rings: a 3-component non-torus link with crossing number $N=6$. This link is a particularly interesting example because the three components are linked, but not pairwise, thus all the component filaments have zero total twist, unlike any of the torus links.
The initial condition for the Borromean rings is obtained using the method for non-torus links discussed in section \ref{sec:FN}, and the associated filament is shown in the first image in both the upper and lower panels of \reffig{fig:Borromean}. 

The evolution shown in the upper panel in \reffig{fig:Borromean} is for a medium with the dimensions $-90\le x,y \le 90$ and $-50\le z\le 80.$ This medium is sufficiently large that bulk evolution ensues. The length of the link initially decreases and although each component has zero total twist, regions of locally large twist rate develop, as evidenced by the colouring of the filament. The filament dynamics becomes increasingly irregular and the link tumbles through the medium until eventually it approaches a boundary of the medium and a filament breaks at this boundary (not shown). Note that up until the point at which the filament breaks at the boundary, the topology is preserved as there are no filament reconnections, despite the irregular motion of each component.

The lower panel in \reffig{fig:Borromean} displays the contrasting evolution using the same initial condition in a slightly smaller and much thinner medium with dimensions 
$-80\le x,y \le 80$ and $0\le z\le 30.$ In this case the motion is regular with the link maintaining its size and a conformation that is not quite as symmetric as the initial condition but is similar. This boundary stabilized link rotates in the same manner found earlier for torus links that sit snuggly at the boundary of the medium, and its stability has been tested over thousands of rotation periods of the underlying vortex that forms the filaments.

The boundary stabilization of vortex rings has recently been demonstrated in experiments on chemical excitable media \cite{ATE,TES} that are modelled by reaction diffusion equations with a very similar structure to the FitzHugh-Nagumo equation. It therefore appears that boundary stabilization is a generic phenomenon that applies to vortex rings, knots and links across a range of excitable media.

\section{Colliding links}
\begin{figure}[h!]
\begin{center}
 \includegraphics[width=0.5\columnwidth]{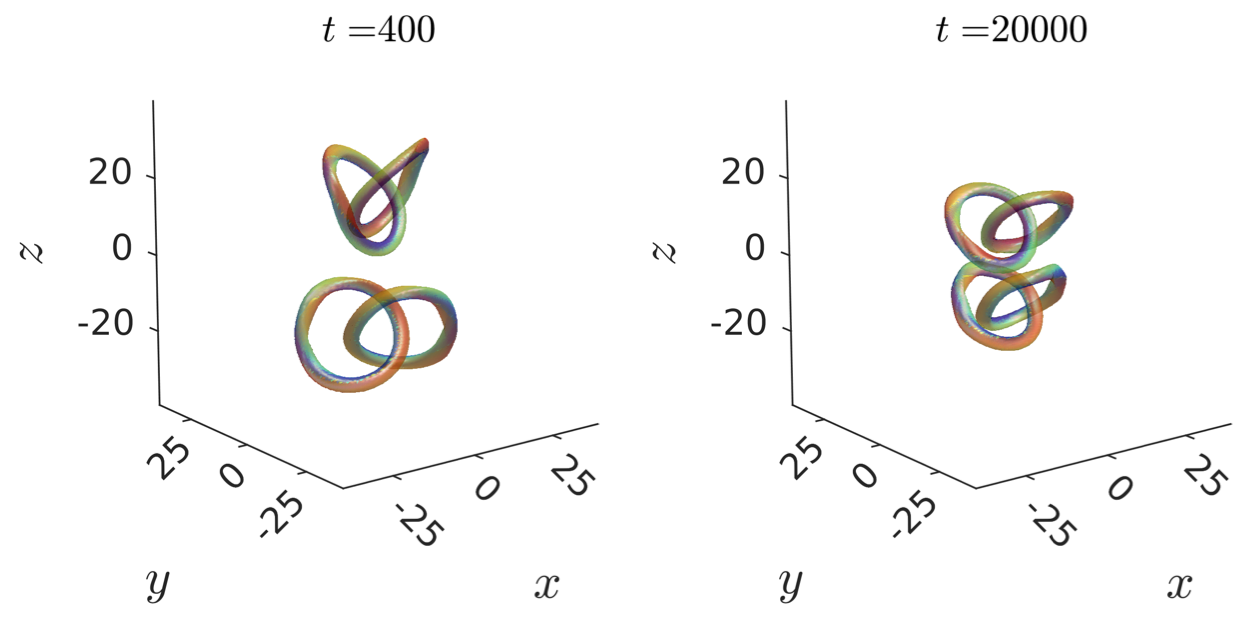}
\end{center}
\caption{The head-on collision of a pair of contra-rotating Hopf links results in a stalemate of mutual blocking.}
\label{fig:no_impact_collision}
\end{figure}

\quad
In this section we consider the collision of a pair of initially well-separated links moving in opposite directions. As we have already seen, a link can snuggly sit indefinitely at a no-flux boundary of the medium and this is equivalent to the collision of a link in the bulk with its mirror image.
The mirror image of a link in the boundary $z=0$ is obtained by making the replacement $(x,y,z)\mapsto(x,y,-z)$, and this reverses the direction of motion of the link. A less symmetric initial condition for a collision can be initiated by applying the transformation $(x,y,z)\mapsto(x,-y,-z)$ to obtain the second link, rather than using the mirror image. This transformation also reverses the direction of motion of the link but yields a pair of contra-rotating links, instead of the co-rotation that follows from using the mirror image. We shall consider contra-rotating links in all the collisions presented in this section, as this is the less symmetric situation.
This simplest example, a pair of Hopf links, is displayed in \reffig{fig:no_impact_collision}, where the first image shows the links just before the collision and the second image demonstrates the stalemate of mutual blocking that persists after thousands of rotation periods of the constituent vortex cores. Despite the contra-rotation, the links sit snuggly together in a comparable manner to a single link at a no-flux boundary. Similar results are obtained for other torus links and knots in the head-on collision of an identical pair.
\begin{figure}[h!]
\begin{center}
 \includegraphics[width=\columnwidth]{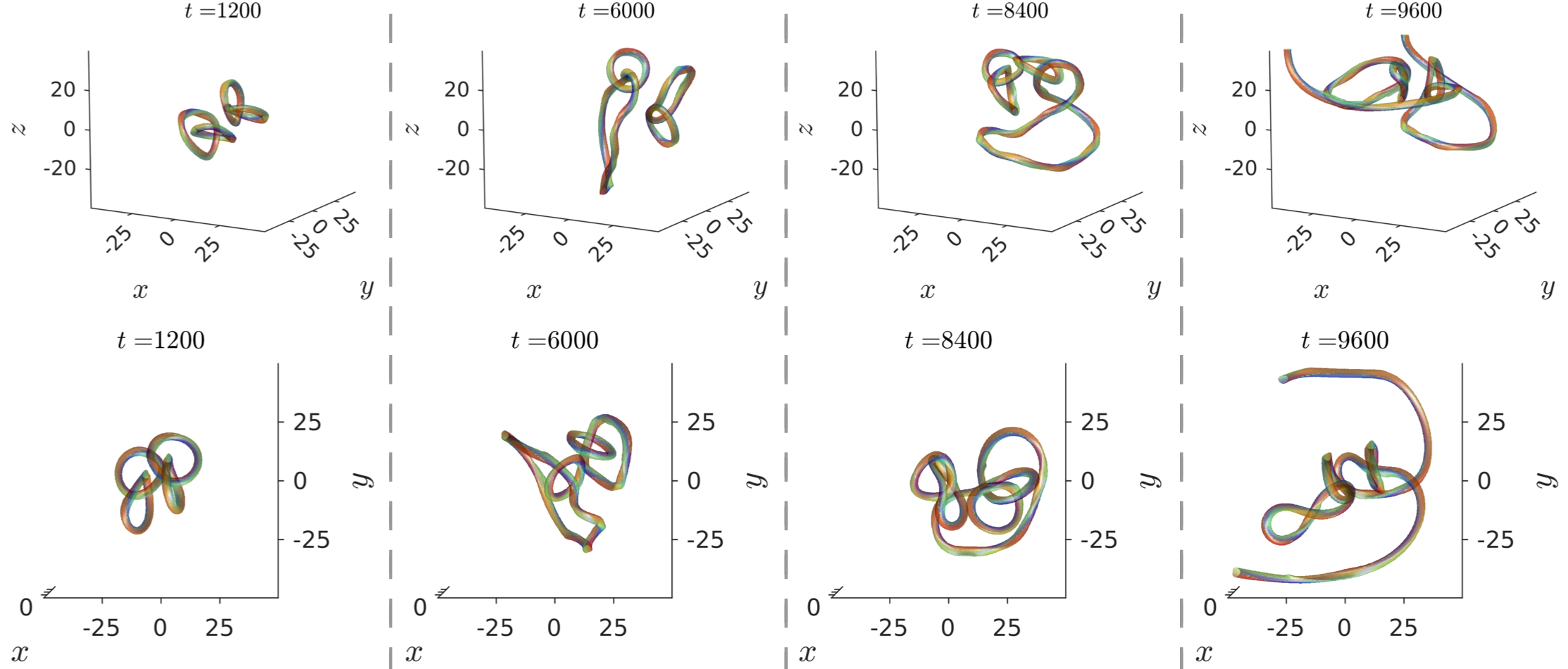}
 \end{center}
 \caption{The collision of a pair of Hopf links with a non-zero impact parameter produces a complex irregular tumbling evolution, where 
 the two links wrestle each other until the filament of one link breaks at the boundary. The upper and lower panels show the filaments at four different times from two different viewing angles.
 }
\label{fig:impact_collision}
\end{figure}
\begin{figure}[h!]
 \includegraphics[width=\columnwidth]{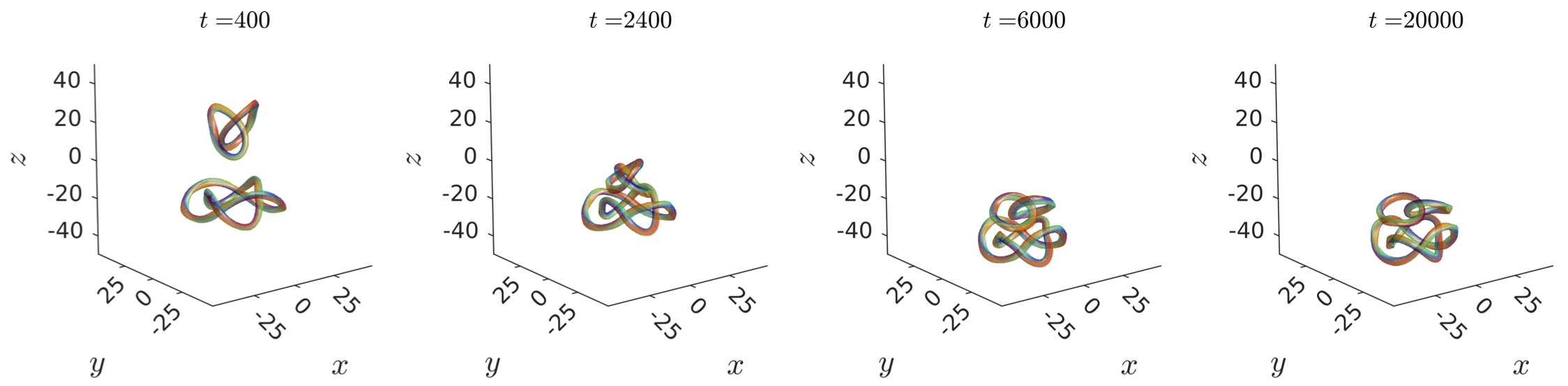}
 \caption{The head-on collision of a Hopf link with Solomon's knot.
   The Hopf link pushes Solomon's knot to the boundary and permanently traps it there.
 }
\label{fig:TL22vsTL42}
\end{figure}
\begin{figure}[h!]
 \includegraphics[width=\columnwidth]{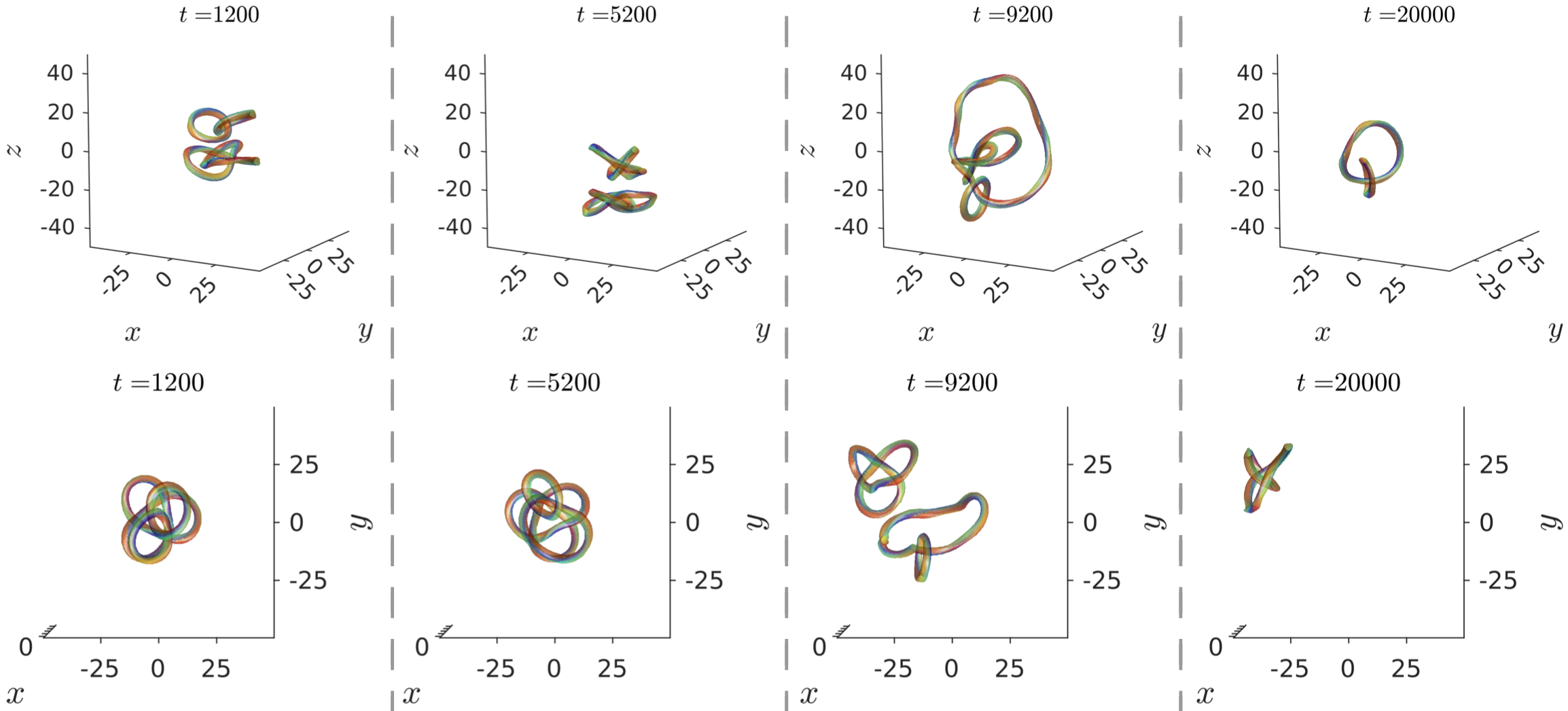}
 \caption{
   A head-on collision of a Hopf link with a trefoil knot. The upper and lower panels show the filaments at four different times from two different viewing angles. The Hopf link pushes the trefoil to the bottom boundary but the asymmetry of the pair allows the trefoil to escape to the side. After an irregular tumbling motion, where the link and knot wrestle each other and change conformation, the knot is pushed into the boundary and its filament breaks, leaving a Hopf link that settles snugly at the boundary.
 }
\label{fig:hopf_trefoil_collision}
\end{figure}

The introduction of a sufficiently large impact parameter for the collision resolves the stalemate of a head-on collision. The upshot is an irregular tumbling dynamics, where the links wrestle each other until one link dominates by pushing the other into the boundary, where the filament breaks. An example is presented in \reffig{fig:impact_collision}, where the upper and lower panels show the filaments at four different times from two different viewing angles. Eventually, the surviving link comes to rest snuggly at the boundary, forming the boundary stabilized Hopf link that we have seen earlier. The precise details of the evolution depend upon the value of the impact parameter and the survival of only a single link is not always the eventual outcome. Another possible outcome that we have observed is the formation of a threaded ring by the breaking of the filament of one link at the boundary, which coexists with the surviving Hopf link settled at a different region of the boundary.

Next we consider the collision of two links of different types.
The example of the head-on collision of a Hopf link with Solomon's knot is displayed in \reffig{fig:TL22vsTL42}. The Hopf link pushes Solomon's knot to the boundary, where the whole structure permanently resides with an unexpected stability. We attribute the dominance of the Hopf link to its speed advantage over Solomon's knot. A similar phenomenon has recently been observed in the dynamics of threaded vortex rings \cite{Sutcliffe:JPhysA:2018}, where smaller rings, with a higher twist rate and frequency of wave emission, travel faster and can push larger rings, even changing their direction of motion.

Finally, we consider the head-on collision between a link and a knot.   
The example with lowest total crossing number is the collision between a Hopf link and a trefoil knot, presented in \reffig{fig:hopf_trefoil_collision}, where the upper and lower panels show the filaments at four different times from two different viewing angles. Once again, the faster moving Hopf link dominates as it pushes the trefoil knot to the bottom boundary. However, even in a head-on collision the asymmetry between the knot and the link  allows the trefoil a means of escape to the side. This generates a period of wrestling between unequal opponents, with irregular tumbling dynamics that results in considerable changes in conformation to both the knot and the link. Eventually, the attempted escape ends in failure as the Hopf link pushes the trefoil knot into a side boundary where its filament breaks. The Hopf link then settles at the boundary, forming the usual boundary stabilized configuration.

\section{Conclusion}\quad
In this paper we have presented the results of parallel numerical computations to systematically investigate the dynamics of torus links in the FitzHugh-Nagumo excitable medium. The results are displayed by using vorticity isosurfaces to identify filaments that are phase-coloured to illustrate twist. Furthermore, by calculating moments of the vorticity, we have analyzed the properties, such as speed, size and rotation frequency, of all torus links with crossing number no greater than 12 and a torus link with crossing number 20 to check asymptotics. We found that these properties fit a simple dependence on crossing number for a fixed number of components of the link.

We have demonstrated that instabilities exist over long timescales for the evolution of links in the bulk, but that these instabilities can be removed by boundary interactions to yield stable torus links, in a manner similar to that found earlier for torus knots \cite{Sutcliffe:PRE:2017}. As an example of a non-torus link, we studied the dynamics of the Borromean rings, which is novel in that the filaments are linked but each has zero total twist. The complexity of the dynamics of bulk evolution of the Borromean rings is beautifully complex, and yet can be regularized by boundary interactions through confinement in a sufficiently tight medium. The fact that filament dynamics is dramatically influenced by the proximity of the medium boundary is of significant relevance to cardiac applications of the FitzHugh-Nagumo model, because in this context the depth of the medium is comparable to a spiral wavelength, so it is certainly evolution within a tight confining medium.

Finally, we considered the collision of initially well-separated links. Head-on collisions lead to regular dynamics with a mutual blocking or trapping of links. However, a non-zero impact parameter produces irregular tumbling dynamics in which links wrestle with each other, while substantially varying their conformation, until one link eventually dominates by pushing the other into the boundary of the medium, where its filament breaks. These results illustrate the remarkable complexity of filament dynamics that occurs on long timescales, with bulk evolution preserving filament topology despite considerable changes in conformation and exotic dynamics. Eventually, regular dynamics prevails once the surviving torus link has settled at the boundary of the medium.

\section*{Acknowledgements}
\noindent  
This work is funded by the
Leverhulme Trust Research Programme Grant RP2013-K-009, SPOCK: Scientific Properties Of Complex Knots.


\begin{thebibliography}{99}

\bibitem{Kelvin}
W. Thomson, 
\textit{Philos. Mag. Ser.} 4, {\bf 34}, 15 (1867).

\bibitem{KI}
  D. Kleckner and W.T.M. Irvine,
  \textit{Nat. Phys.} {\bf 9}, 253 (2013).

\bibitem{Den}
  M.R. Dennis, R.P. King, B. Jack, K. O'Holleran and M.J. Padgett,
 \textit{Nat. Phys.} {\bf 6}, 118 (2010).

 \bibitem{AS2}
  P.J. Ackerman and  I.I. Smalyukh, 
  \textit{Phys. Rev. X} {\bf 7}, 011006 (2012).
  
\bibitem{FN} L. Faddeev and A.J. Niemi,
  \textit{Nature} {\bf 387}, 58 (1997).

\bibitem{Bar} D. Proment, M. Onorato and C.F. Barenghi,
  \textit{Phys. Rev. E} {\bf 85}, 036306 (2012). 
  
\bibitem{Sut} P.M. Sutcliffe,
  \textit{Phys. Rev. Lett.} {\bf 118}, 247203 (2017).  
  
\bibitem{FH}
   R. FitzHugh,
   \textit{Biophys. J.} {\bf 1}, 445 (1961).

 \bibitem{Nag}  
   J.S. Nagumo, S. Arimoto and S. Yoshizawa,
   \textit{Proc. IRE.} {\bf 50}, 2061 (1962).

 \bibitem{Kog} B.Y. Kogan, \textit{Introduction to computational cardiology},   
   Springer, 2010.
   
 \bibitem{Izh} E.M. Izhikevich, 
   \textit{Dynamical Systems in Neuroscience: The Geometry of Excitability and Bursting}, The MIT Press, 2007.

 \bibitem{Winfree:Nature:1984}
   A.T. Winfree and S.H. Strogatz,
  \textit{Nature} {\bf 311}, 611 (1984).

\bibitem{Sutcliffe:PRE:2003}
P.M. Sutcliffe and A.T. Winfree,
\textit{Phys. Rev. E} {\bf 68}, 016218 (2003).

\bibitem{Sutcliffe:PRL:2016}
 F. Maucher and P.M. Sutcliffe,
 \textit{Phys. Rev. Lett.} {\bf 116}, 178101 (2016).

\bibitem{Sutcliffe:PRE:2017}
 F. Maucher and P.M. Sutcliffe,
 \textit{Phys. Rev. E} {\bf 96}, 012218 (2017).

\bibitem{winfree:physD:1985}
  A.T. Winfree, E.M. Winfree and H. Seifert, \textit{Physica D} {\bf 17}, 109 (1985).

\bibitem{Winfree:SIAM:1990}
A.T. Winfree,
  \textit{SIAM Review} {\bf 32}, 1 (1990).

\bibitem{Winfree:Nature:1994}
  A.T. Winfree,
  \textit{Nature} {\bf 371}, 233 (1994).

 \bibitem{Sutcliffe:JPhysA:2018} 
F. Maucher and P.M. Sutcliffe,
\textit{J. Phys. A} {\bf 51}, 055102 (2018).

\bibitem{Winfree} A.T. Winfree, \textit{The Geometry of Biological Time},
  Springer-Verlag, 2001.

\bibitem{Win3}
 A.T. Winfree,
 \textit{Physica D} {\bf 84}, 126 (1995).

\bibitem{Winfree:book:2002}
  A.T. Winfree, in {\em Nonlinear Dynamics and Chaos: Where do we go from here?}, edited by S.J. Hogan {\em et al.}, IOP, London, 2002.

\bibitem{Milnor}
  J. Milnor,
  \textit{Singular Points of Complex Hypersurfaces},
  Annals of Mathematics Studies, Princeton University Press, 1969.

\bibitem{Keener:PhysicaD:1988}
  J.P. Keener,
  \textit{Physica D} {\bf 31}, 269 (1988).

\bibitem{Biktashev:1994}
V.N. Biktashev, A.V. Holden and H. Zang,
  \textit{Philos. Trans. R. Soc. London} {\bf A347}, 611 (1994). 
    
\bibitem{ATE}
A. Azhand, J.F. Totz and H. Engel,
\textit{EPL} {\bf 108}, 10004 (2014).

\bibitem{TES}
  J.F. Totz, H. Engel and O. Steinbock,
  \textit{New J. Phys.} {\bf 17}, 093043 (2015).
  
\end{thebibliography}
\end{document}